\documentclass{ws-ijgmmp}

\usepackage{xcolor}
\usepackage{physics}
\usepackage{MnSymbol}
\usepackage{dsfont}

\usepackage{cite}

\begin{document}

\markboth{Bernard, P.-A., Parez, G., Vinet, L.}
{Distinctive features of inhomogeneous spin chains}

%
\catchline{}{}{}{}{}
%

\title{Distinctive features of inhomogeneous spin chains}

\author{PIERRE-ANTOINE BERNARD}

\address{Centre de Recherches Mathématiques (CRM)\\ Université de Montréal, C.P. 6128, Succursale Centre-ville\\
Montréal, Québec,  H3C 3J7, Canada\\
\email{bernardpierreantoine@outlook.com} }

\author{GILLES PAREZ}
\address{Laboratoire d’Annecy-le-Vieux de Physique Théorique (LAPTh) \\ CNRS, Université Savoie Mont Blanc \\ 74940 Annecy, France \\ \email{parez@lapth.cnrs.fr} }

\author{LUC VINET}

\address{IVADO and Centre de Recherches Mathématiques (CRM)\\ Université de Montréal, C.P. 6128, Succursale Centre-ville\\
Montréal, Québec,  H3C 3J7, Canada\\
\email{luc.vinet@umontreal.ca} }

\maketitle

\begin{history}
\received{(Day Month Year)}
\revised{(Day Month Year)}
\end{history}

\begin{abstract}
This review presents recent developments in the study of inhomogeneous XX spin chains, highlighting results on perfect state transfer, out-of-equilibrium stationary dynamics in open systems, and entanglement and correlations in ground states. We discuss the conditions on couplings that enable perfect state transfer, examine how heat currents scale when the chains are coupled to thermal baths, explore the role of tridiagonal matrices in approximating the entanglement Hamiltonian and investigate bulk and boundary entanglement negativity and correlation decay. These findings underscore some of the distinctive physical behavior of inhomogeneous spin chains and their potential applications in quantum information and thermal transport.
\end{abstract}


\section{Introduction}

Spin chains offer a fruitful test-bed to study entanglement and non-equilibrium steady states in quantum many-body systems. While most of these analyses have focused on homogeneous models with uniform couplings between neighboring sites, realistic physical systems  often contain defects and inhomogeneities. This can be addressed by considering spin chains with non-homogeneous couplings and it turns out that the corresponding models exhibit a range of intriguing and specific properties, including perfect state transfer (PST) and fractional revival (FR). Nevertheless, inhomogeneous chains remain comparatively much less explored relative to their homogeneous counterparts.

Recent works realized with collaborators have initiated a systematic examination of inhomogeneous XX spin chains. This paper reviews key findings from these studies. In Sec.~\ref{sec:2}, we introduce the Hamiltonian of the generic non-uniform XX spin chain and present its connection with a free fermion model. We further outline the standard approach to diagonalizing this Hamiltonian and highlight the links to the theory of orthogonal polynomials. Section~\ref{sec:3} reviews the perfect state transfer protocol, detailing the special role of inhomogeneous chains in this context. Section~\ref{sec:4} addresses the behavior of these chains in open settings when coupled to thermal baths, with a focus on the scaling of heat currents and their relation to PST. Sections~\ref{sec:5}, \ref{sec:6} and \ref{sec:7} explore entanglement and correlation properties in these models, emphasizing a simple tridiagonal matrix that commutes with the truncated correlation matrix and approximates the entanglement Hamiltonian, as well as unusual boundary correlation decay.

\section{Inhomogeneous XX chains}\label{sec:2}

The Hamiltonian of an inhomogeneous XX spin chain composed of $N+1$ spins is given by  
\begin{equation}\label{eq:hamil}
    \mathcal{H} = -\frac{1}{2} \sum_{\ell =0}^{N-1} J_\ell (\sigma_\ell^x \sigma_{\ell + 1}^x + \sigma_\ell^y \sigma_{\ell + 1}^y ) + \frac{1}{2} \sum_{\ell = 0}^N B_\ell ( 1+ \sigma_\ell^z),
\end{equation}
with $J_\ell$ the coupling between the spins $\ell$ and $\ell + 1$, and $B_\ell$ the amplitude of the magnetic field at site $\ell$. This operator acts on the Hilbert space $(\mathbb{C}^2)^{\otimes (N+1)}$ of $N+1$ spins and admits a conserved charge corresponding to the $z$-component of the total spin operator, i.e.
\begin{equation}
    S^z = \sum_{\ell = 0}^N \sigma_\ell^z, \quad \quad [\mathcal{H}, S^z] = 0.
\end{equation}
It follows that the Hilbert space of states decomposes into subspaces labeled by the number of spins that are up and on which the action of $ \mathcal{H} $ is closed. This rotation symmetry plays a key role in the diagonalization of this Hamiltonian and the study of perfect state transfer as will be seen in Sec.~\ref{sec:3}.
\\ 

\noindent \textbf{Connection with free fermions.} The Hamiltonian $ \mathcal{H} $ is mapped to a free-fermion model using the Jordan-Wigner transformation, which expresses the local spin operators in terms of fermionic creation and annihilation operators,
\begin{equation}
    c_\ell^\dagger = \sigma_0^z \sigma_1^z \dots \sigma_{\ell -1}^z \sigma_\ell^+, \quad c_\ell = \sigma_0^z \sigma_1^z \dots \sigma_{\ell -1}^z \sigma_\ell^-,
\end{equation}
where $\sigma_\ell^\pm = \frac{1}{2}(\sigma_\ell^x \pm i \sigma_\ell^y)$. These operators satisfy the canonical anticommutation relations
\begin{equation}\label{eq:antic}
    \{c_k, c_\ell \} = \{c_k^\dagger, c_\ell^\dagger \} = 0, \quad \{c_k, c_\ell^\dagger \} = \delta_{k \ell}.
\end{equation}
Using this transformation, the Hamiltonian $ \mathcal{H} $ is rewritten as
\begin{equation}\label{eq:defHamil}
    \mathcal{H} = \sum_{k,\ell = 0}^N \Lambda_{k\ell} c_k^\dagger c_\ell,
\end{equation}
where $\Lambda$ is an $(N+1) \times (N+1)$ tridiagonal matrix formed out of the hopping coefficients and the magnetic fields,
\begin{equation}
    \Lambda = \begin{pmatrix}
        B_0 & J_0 &  &  &  &  \\
        J_0 & B_1 & J_1 &  &  &  \\
         & J_1 & B_2 & J_2 &  &  \\
        &  & \ddots & \ddots & \ddots  &  \\
         &   &    & J_{N-2} & B_{N-1} & J_{N-1}  \\
          &   &     &   & J_{N-1} & B_{N}  \\
    \end{pmatrix}.
\end{equation}

\noindent \textbf{Diagonalization of $\mathcal{H}$. } The diagonalization of the free-fermion Hamiltonian of Eq.~\eqref{eq:defHamil} proceeds in the following manner. Let $|0\rrangle$ denote the vacuum state annihilated by all operators $c_\ell$, i.e.
\begin{equation}
    c_\ell |0\rrangle = 0, \quad \forall \ell = 0,1,\dots,N.
\end{equation}
It is straightforward to check that the action of $\mathcal{H}$ on the $N+1$ dimensional subspace of $1$-excitation states, $ c_\ell^\dagger |0\rrangle$ with $\ell= 0, 1, \dots, N$, is closed and corresponds to the tridiagonal action of $\Lambda$ on  column vectors $\ket{\ell} \in \mathbb{C}^{N+1}$, upon the identifications $\mathcal{H}|_{1\text{-excitation}} \leftrightarrow \Lambda$ and $\ket{\ell} 
 \leftrightarrow c_\ell^\dagger|0\rrangle$,
\begin{equation}
    \mathcal{H} c_\ell^\dagger |0\rrangle = J_{\ell} c_{\ell+1}^\dagger |0\rrangle + B_\ell c_{\ell}^\dagger |0\rrangle+ J_{\ell-1} c_{\ell-1}^\dagger |0\rrangle.
\end{equation}
 The construction of the $1$-excitation stationary states thus amounts to solving the eigenvalue problem for the matrix $\Lambda$, which reads
\begin{equation}
    \Lambda \ket{\omega_k} = \omega_k \ket{\omega_k}, \quad \ket{\omega_k} = \sum_{\ell = 0}^N \phi_\ell(\omega_k) \ket{\ell}.
\end{equation}
By looking at $\bra{\ell} \Lambda \ket{\omega_k} = \left(\bra{\ell} \Lambda^T\right)\ket{\omega_k}$, this spectral problem is seen to be equivalent to solving the following three-term recurrence relation,
\begin{equation}\label{eq:3rec}
    \omega_k \phi_\ell(\omega_k) = J_\ell \phi_{\ell + 1}(\omega_k) + B_\ell \phi_{\ell}(\omega_k) + J_{\ell -1} \phi_{\ell - 1}(\omega_k).
\end{equation}
Owing to Favard's theorem, the $1$-excitation wavefunctions $\phi_\ell$ can thus be expressed in terms of orthogonal polynomials $\chi_\ell$ of order $\ell$ in $\omega_k$, i.e.
\begin{equation}
\phi_\ell(\omega_k) = \phi_0(\omega_k) \chi_\ell(\omega_k), \quad \sum_{k=0}^N \phi_0^2(\omega_k) \chi_i(\omega_k) \chi_j(\omega_k) = \delta_{ij},
\end{equation}
with $\chi_0=1$ and $\phi_0^2(\omega_k)$ their weight function. The wavefunctions $\phi_\ell(\omega_k)$ allow for the definition of new creation and annihilation operators $\tilde{c}_k^\dagger$ and $\tilde{c}_k$ which satisfy the canonical fermionic relations \eqref{eq:antic}, and in terms of which the Hamiltonian is diagonal,
\begin{equation}
    \mathcal{H} = \sum_{k = 0}^N \omega_k \Tilde{c}^\dagger_k \Tilde{c}_k , \quad  \Tilde{c}_k = \sum_{\ell = 0}^N \phi_\ell(\omega_k) c_\ell.
\end{equation}
An $m$-excitation stationary state is then obtained by acting with $m$ distinct creation operators $\tilde{c}_k^\dagger$ upon the vacuum state. Using the canonical fermionic anticommutation relations, one finds that these states are orthonormal and that the action of $\mathcal{H}$ on them is diagonal,
\begin{equation}
     \mathcal{H} \Tilde{c}_{k_1}^\dagger \Tilde{c}_{k_2}^\dagger \dots \Tilde{c}_{k_m}^\dagger |0 \rrangle = \left(\sum_{i = 1}^m \omega_{k_i}\right) \Tilde{c}_{k_1}^\dagger \Tilde{c}_{k_2}^\dagger \dots \Tilde{c}_{k_m}^\dagger |0 \rrangle.
\end{equation}
The fact that they provide a complete basis follows from their orthogonality and a cardinality argument. In the upcoming sections, these stationary states will be denoted as
\begin{equation}\label{eq:stat_state}
    |{\vec n}\rrangle = \left(\tilde{c}_0^\dagger\right)^{n_0}\left(\tilde{c}_1^\dagger\right)^{n_1}\dots \left(\tilde{c}_N^\dagger\right)^{n_N} | 0 \rrangle,
\end{equation}
where $\vec{n} = (n_0, n_1, \dots, n_{N})$ is an $N+1$ dimensional vector with entries $n_i \in \{0,1\}$.

\section{Perfect state transfer}\label{sec:3}

Quantum communication channels are vital to many key protocols, with quantum teleportation being a notable example. In \cite{bose2003quantum}, Bose first proposed utilizing the intrinsic dynamics of spin chains to achieve this with minimal external intervention. His and subsequent studies have shown that end-to-end qubit transfer with perfect fidelity does not occur in homogeneous spin chains with more than four sites, but can be achieved in inhomogeneous ones with carefully engineered couplings \cite{christandl2004perfect, kay2010perfect, vinet2012construct}. Let us revisit the conditions on $J_\ell$ and $B_\ell$ required for a chain to exhibit \textit{perfect state transfer} (PST). \\

\noindent \textbf{Conditions for PST.} Perfect state transfer is characterized within the $1$-excitation subspace spanned by the vectors $\ket{\ell}$, $\ell = 0, 1, \dots, N$, defined above. A chain is said to exhibit PST at time $\tau$ when the state $ \ket{0} $, representing an excitation localized at site $ 0 $, is mapped by the unitary time evolution $e^{-i \tau \Lambda}$ to the state $ \ket{N} $, representing an excitation localized at site $ N $ at the opposite end of the chain:
\begin{equation}\label{eq:condPST0}
    e^{-i\tau\Lambda} \ket{0} = e^{i \phi} \ket{N}.
\end{equation}
An analysis based on the overlaps of the states on both sides of Eq.~\eqref{eq:condPST0} with the eigenbasis of $\Lambda$ shows that this occurs if and only if the following two conditions hold:
\begin{equation}\label{eq:condPST1}
\chi_N(\omega_k) = (-1)^{N+k},
\end{equation}
\begin{equation}\label{eq:condPST2}
\omega_{k+1} -\omega_k = \frac{\pi}{\tau} M_k, \quad M_k \in \{1, 3, 5, \dots \},
\end{equation}
with $\chi_N$ and $\omega_k$ the orthogonal polynomials and eigenvalues introduced in Sec.~\ref{sec:2}. From standard arguments coming from the study of orthogonal polynomials, it can be shown that condition \eqref{eq:condPST1} is equivalent to the single-excitation Hamiltonian $\Lambda$ being mirror-symmetric, i.e.
\begin{equation}\label{eq:condPST3}
    B_{N-n} = B_n, \quad J_{N+1-n} = J_n.
\end{equation}
Given a spectrum that satisfies \eqref{eq:condPST2}, the mirror-symmetric coefficients $ J_\ell $ and $ B_\ell $ of a spin chain exhibiting perfect state transfer can be determined by solving an inverse spectral problem \cite{vinet2012construct}. This proceeds as illustrated below using Euclidean division, which thus gives a constructive approach to the identification of spin chains exhibiting PST.\\

\noindent \textbf{Inverse spectral problem.} The tridiagonal mirror-symmetric matrix $ \Lambda $, whose eigenvalues correspond to a given set $ \{\omega_k \,|\, k = 0,1,\dots N\}$ satisfying \eqref{eq:condPST2}, is indeed determined efficiently using a straightforward application of the Euclidean algorithm. Let $P_\ell$ denote the monic polynomial obtained by dividing the polynomial $\chi_\ell$ of order $\ell$ by its leading coefficient. The spectrum $ \{\omega_k \,|\, k = 0,1,\dots N\} $ is encoded in $P_{N+1}$, the characteristic polynomial of $\Lambda$:
\begin{equation}
    P_{N+1}(x) = (x - \omega_0)(x - \omega_1)\dots(x-\omega_{N}).
\end{equation}
Condition \eqref{eq:condPST1} allows the construction of the monic polynomial $P_N$ through Lagrange interpolation. Given the polynomials $P_{N+1}$ and $P_N$, one can apply Euclidean division and identify the remainder as a multiple of $P_{N-1}$. This procedure can be repeated to obtain the remaining polynomials $P_\ell$. The coefficients $J_\ell$ and $B_\ell$ are then determined by their correspondence with the coefficients in the recurrence relation of the monic polynomial $P_\ell$, which themselves arise as the coefficients obtained during the Euclidean division of $P_{\ell+1}$ by $P_\ell$. The three term recurrence of these polynomials follows from \eqref{eq:3rec} and is given by
\begin{equation}
    P_{\ell+1}(x) + (B_\ell -x)P_{\ell}(x) + J_{\ell -1}^2 P_{\ell-1}(x) =0,
\end{equation}
where one identifies respectively $Q(x) = (x - B_\ell)$ and $R(x) = - J_{\ell -1}^2 P_{\ell-1}(x)$ as the quotient and remainder of the division of $P_{\ell+1}(x)$ by $P_{\ell}(x)$. \\

\noindent \textbf{Examples.} Let us consider the linear case $\omega_k = k$, which satisfies \eqref{eq:condPST2} with $M_k = 1$ and $\tau = \pi$. The application as described above of the Euclidean algorithm leads to
\begin{equation}
    J_n = \frac{1}{2} \sqrt{(n+1)(N-n)}, \quad B_n = N/2.
\end{equation}
These couplings correspond to the special case $p=1/2$ of the coefficients
\begin{equation}\label{eq:coeff_kraw}
    J_n = \sqrt{p(1-p)} \sqrt{(n+1)(N-n)}, \quad B_n  = p (N-n) + (1-p) n
\end{equation}
where $p \in [0,1]$. It is well known that the wavefunctions solving Eq.~\eqref{eq:3rec} with those $J_n$ and $B_n$ are expressed as follows in terms of Krawtchouk polynomials $K_n(k;p,N)$ \cite{koekoek}:
\begin{equation}
    \phi_n(\omega_k) = (-1)^n\sqrt{\binom{N}{n}\binom{N}{k} p^{k+n}(1-p)^{N-k-n}} K_n(k;p,N).
\end{equation}
The associated chain is referred to as the \textit{Krawtchouk chain}, and is known to exhibit perfect state transfer at 
$p= 1/2$, when mirror symmetry is enforced. Another natural example to consider is the homogeneous chain defined by the couplings
\begin{equation}
    J_n = 1, \quad B_n = 0.
\end{equation}
Its $1$-excitation spectrum and wavefunctions are \cite{christandl2004perfect, crampe2019free}
\begin{equation}
    \omega_k = 2\cos\left(\frac{\pi (k+1)}{N+2}\right), \quad     \phi_n(\omega_k) = \sqrt{\frac{2}{N+2}} {\sin} \left(\frac{\pi (k+1)}{N+2}\right) U_n(\omega_k),
\end{equation}
where $U_n(\omega_k)$ are the Chebyshev orthogonal polynomials of the second kind. It is apparent that for this example $\omega_{k+1} - \omega_k \neq \frac{\pi}{\tau} M_k$, and as a consequence, this chain does not exhibit no PST.

These two examples highlight the necessity of inhomogeneous couplings to realize perfect state transfer. While this may seem at first glance to be an obstacle for the experimental realization of the protocol, good control over the coupling strengths is now possible within quantum devices. PST with Krawtchouk couplings was notably realized with photonic qubits and waveguides in \cite{chapman2016experimental}.

\section{Non-equilibrium steady-states of open inhomogeneous chains}\label{sec:4}

While the study of PST presented above was concerned with the intrinsic dynamics of a spin chain isolated from its environment, the end-to-end transport of spin can also be investigated in open settings by characterizing heat and spin currents in non-equilibrium steady states. This is of notable interest for modeling the magnetic contribution to thermal conductivity in certain materials \cite{hlubek2010ballistic,pan2022unambiguous,sologubenko2007thermal}. Since these materials may exhibit defects leading to inhomogeneous spin couplings, inspecting the heat and spin currents in non-uniform spin chains is quite relevant. We will now describe recent results obtained in \cite{benatti2022stationary,  benatti2021exact, bernard2024currents} concerning the generic spin chain with Hamiltonian given in Eq.~\eqref{eq:hamil}, when coupled to bosonic heat baths at its ends.\\

\noindent \textbf{Open model.}  We take the Hamiltonians of the baths to be given by
\begin{equation}
\mathcal{H}_\alpha=\int_0^\infty\!\mathrm{d}\nu\ \nu \mathfrak{b}_\alpha^\dagger(\nu)\mathfrak{b}_\alpha(\nu),\qquad \alpha\in\{0,N\},
\end{equation}
where $\mathfrak{b}_\alpha(\nu)$ and $\mathfrak{b}_\alpha^\dagger(\nu)$ are bosonic creation and annihilation operators that satisfy the canonical commutation relations:
\begin{equation} \left[\mathfrak{b}_\alpha^\dagger(\nu),\mathfrak{b}_\beta(\nu')\right]=\delta_{\alpha\beta}\delta(\nu-\nu').
\end{equation}
The states of the two baths are expressed in terms of their respective Hamiltonians $\mathcal{H}_\alpha$ and inverse temperatures $\beta_\alpha = 1/T_\alpha$,
\begin{equation}
\rho_\alpha=\frac{e^{-\beta_\alpha \mathcal{H}_\alpha}}{\mathrm{Tr}\left(e^{-\beta_\alpha \mathcal{H}_\alpha}\right)},\qquad \alpha\in\{0,N\}.
\end{equation}
The interaction between the baths and the chain is modeled by the following Hamiltonian:
\begin{equation}\label{eq:interaction-hamiltonian}
    \mathcal{H}_I=\sum_{\alpha\in\{0,N\}}\left(\sigma^+_\alpha\mathfrak{B}_\alpha+\sigma^-_\alpha\mathfrak{B}^\dagger_\alpha\right),
\end{equation}
where 
\begin{equation}
    \mathfrak{B}_\alpha=\int_0^\infty\mathrm d\nu\ h_\alpha(\nu)\mathfrak{b}_\alpha(\nu),
\end{equation}
with $h_\alpha(\nu)$ being a real, suitable smearing function that serves to introduce a cut-off. The total Hamiltonian of the spin chain and the two baths is thus
\begin{equation}
\mathcal{H}_\mathrm{tot}=\mathcal{H}+\lambda \mathcal{H}_I+\mathcal{H}_0 + \mathcal{H}_N.
\end{equation}
Assuming a weak coupling 
$\lambda$, one can derive a Lindblad master equation using a global approach based on the standard Born-Markov and secular approximations~\cite{breuer2002theory}. This leads to the following dynamical equation for the density matrix $\rho(t)$ of the spin chain,
\begin{equation}\label{eq:Lindblad1}
    \frac{\mathrm d}{\mathrm dt}\rho(t)=-i\left[\mathcal{H}+\lambda^2\mathcal{H}_{LS},\rho(t)\right]+\mathds{D}[\rho(t)]=\mathds{L}[\rho],
\end{equation}
 where $\mathcal{H}_{LS}$ is a Lamb-shift Hamiltonian correcting the energy levels. An expression for $\mathcal{H}_{LS}$ in terms of quadratic operators $\tilde{c}_k^\dagger \tilde{c}_k$ is provided in \cite{bernard2024currents} and is omitted here as it does not play a role in study of spin and heat currents. The dissipative part~$\mathds{D}$ in Eq.~\eqref{eq:Lindblad1} is given by
\begin{equation} \label{eq:clean-diss}
    \mathds{D}[\rho]=\lambda^2\sum_{k=0}^N \left( d_k\left(\tilde{c}_k^\dagger\rho(t)\tilde{c}_k-\frac{1}{2} \left\{\rho(t),\tilde{c}_k \tilde{c}_k^\dagger\right\}\right)\right. +\left.\tilde{d}_k\left(\tilde{c}_k\rho(t)\tilde{c}_k^\dagger-\frac{1}{2} \left\{\rho(t),\tilde{c}_k^\dagger \tilde{c}_k\right\}\right)\right).
\end{equation}
The coefficients $d_k$ and $\tilde{d}_k$ that arise in $\mathds{D}$ are expressed in terms of the smearing functions, the particle distributions over the energy, and the $1$-excitation wavefunctions $\phi_0(\omega_k)$ and $\phi_N(\omega_k)$ of the spin chain,
\begin{equation}\label{eq:def-dk}
    \begin{split}
        d_k&=\sum_{\alpha\in\{0,N\}}2\pi \phi_\alpha(\omega_k)^2h_\alpha(\omega_k)^2(n_\alpha(\omega_k)+1), \\
        \tilde{d}_k&=\sum_{\alpha\in\{0,N\}}2\pi \phi_\alpha(\omega_k)^2h_\alpha(\omega_k)^2n_\alpha(\omega_k),
    \end{split}
\end{equation}
with $n_\alpha(\omega)=\frac{1}{e^{\beta_\alpha \omega}-1}$ the Bose–Einstein distribution. \\

\noindent \textbf{Steady state.} The non-equilibrium steady state 
$\rho_\infty$ is defined as the time-invariant state attained by the spin chain after prolonged exposure to the dissipative effects of the baths. The Lindblad equation \eqref{eq:Lindblad1} requires that this state satisfies
\begin{equation}
    \mathds{L}[\rho_\infty] = 0.
\end{equation}
It was observed in \cite{benatti2021exact} that a solution to this equation for the case of an homogeneous spin chain can be found using the following ansatz:
\begin{equation}
\rho_\infty=\sum_{\vec n}\lambda_{\vec n}|{\vec n}\rrangle \llangle{\vec n}|,
\end{equation}
where $|\vec{n}\rrangle$ corresponds to the stationary state as defined in Eq.~\eqref{eq:stat_state}. In \cite{bernard2024currents}, it was shown that this also leads to a solution for the general inhomogeneous case. The solution is given by
\begin{equation}
    \lambda_{\vec n}=\frac{y_{\vec n}}{\prod_{k=0}^N(d_k+\tilde{d}_k)},
\end{equation}
where
\begin{equation}
    y_{\vec n}=\prod_{k=0}^N\left( n_k\tilde d_k+(1-n_k)d_k\right).
\end{equation}

\noindent \textbf{Heat currents. } Since we are considering a global master equation derived using the secular approximation, the first law of thermodynamics holds \cite{soret2022thermodynamic}. Additionally, the time-independence of the Hamiltonian of the spin chain 
$\mathcal{H}$ ensures that the heat flow 
$\mathfrak{h}(t)$ in the system corresponds to the variation in its internal energy, i.e.
\begin{equation}
    \mathfrak{h}(t) = \dv{}{t} \text{Tr}\left(\mathcal{H} \rho(t)\right) = \text{Tr}\left(\tilde{\mathds{L}}[\mathcal{H}]\rho(t)\right),
\end{equation}
where the adjoint Lindbladian $\tilde{\mathds{L}}$ is defined by exploiting the cyclicity of the trace. Its dissipative part can be decomposed as a sum of contributions from the right and left baths. The heat flow can thus be decomposed in a similar way into the currents $\mathfrak{h}_L$ and $\mathfrak{h}_R$, which originate from the left and right baths, respectively:
\begin{equation}
    \mathfrak{h}(t) = \mathfrak{h}_L(t) + \mathfrak{h}_R(t).
\end{equation}
In the case of the time-independent non-equilibrium steady state $\rho_\infty$, the associated currents do not depend on time and satisfy $\mathfrak{h}_L = -\mathfrak{h}_R$. They can be computed straightforwardly from the explicit expression of $\rho_\infty$, which yields
\begin{equation}\label{eq:general-left-heat}
\mathfrak{h}_{L}=2\pi h^2\lambda^2  \sum_k \frac{\omega_k \phi_0(\omega_k )^2\phi_N(\omega_k )^2(n_0(\omega_k )-n_N(\omega_k ))}{\phi_0(\omega_k )^2(2n_0(\omega_k )+1)+\phi_N(\omega_k )^2(2n_N(\omega_k  )+1)},
\end{equation}
 where $h$ is the approximate value of the smearing functions $h_\alpha(\nu)$ evaluted on the spectrum $\{\omega_k \, | \, k = 0,1,\dots, N\}$. The approximation of $h_\alpha(\nu)$ as a constant is justified since these functions are used to introduce a cutoff and are assumed to vary slowly at the energy scale of the spin chain.

In the case of a spin chain with mirror-symmetric couplings and magnetic fields, one can use the symmetry $\phi_0(\omega_k)^2 =  \phi_N(\omega_k)^2$ (as in the case of the Krawtchouk chain with $p=1/2$) to further simplify the expression for the currents, which is then expressed in terms of the $1$-excitation Hamiltonian $\Lambda$,
 \begin{equation}\label{eq:heat-mir}
     \mathfrak{h}_{L}=\pi \lambda^2\bra{0}\Lambda\frac{\sinh\left(\frac{\beta_N-\beta_0}{2}(\Lambda)\right)}{\sinh\left(\frac{\beta_0+\beta_N}{2}(\Lambda)\right)}\ket{0}.
 \end{equation}
Equations \eqref{eq:general-left-heat} and \eqref{eq:heat-mir} characterize the influence of the bath temperatures and the chain length on the heat currents. In the small temperature gradient regime, $|T_{0} - T_N| \ll 1$, these equations align with Fourier's law, $q = \kappa \nabla T$, where $q = \mathfrak{h}_{L}$ and $\nabla T = (T_N - T_0)/N$. This correspondence allows the identification of the magnetic heat conductivity $\kappa$ and reveals cases where $\kappa$ depends on $N$ and thus becomes anomalous. For instance, in the case of a mirror-symmetric chain, one finds that
\begin{align}
    \label{eq:cond2} \kappa &\sim \frac{ \pi \lambda^2 h^2 N }{T^2} \bra{0} {\Lambda^2} e^{- \frac{\Lambda}{T}}\ket{0}, \quad & T \ll 1,\\
    \label{eq:high-temp-reg} \kappa &\sim \frac{ \pi \lambda^2 h^2 N }{2T} B_0, \quad & T \gg 1,
\end{align}
where $T = (T_0 + T_N)/2$. Equation \eqref{eq:cond2} describes a conductivity $\kappa$ associated with ballistic, diffusive, or subdiffusive transport, depending on how the details of the wavefunctions $\phi_0(\omega_k) = \pm \phi_N(\omega_k)$ influence its dependency on $N$. In contrast, Eq.~\eqref{eq:high-temp-reg} gives a conductivity proportional to $N$ and is therefore associated with ballistic transport for any mirror-symmetric chain, at high temperatures. The role of symmetry in this phenomenon is crucial, as indeed demonstrated in \cite{bernard2024currents}, ballistic transport does not to hold in general for chains lacking mirror symmetry in their couplings. Interestingly, small asymmetric perturbations of mirror symmetric couplings have been shown in multiple cases to induce a transition from ballistic to subdiffusive transport at high temperatures. 

Since mirror symmetry is essential for perfect state transfer, these observations demonstrate that the ability of an inhomogeneous spin chain to support spin and heat currents when coupled to a thermal bath is influenced by features that similarly govern its capacity to exhibit PST. It remains an open question whether the spectral condition \eqref{eq:condPST2} could be linked to any thermodynamic properties of the spin chain, in a way analogous to how mirror symmetry appears to be connected to ballistic transport at high temperatures.

\section{Entanglement entropy}\label{sec:5}

Given the significant role of entanglement in quantum algorithms and in the understanding of quantum many-body systems, an intriguing question to explore is how entanglement is influenced by varying couplings in inhomogeneous spin chains. One approach to address this is to analyze the entanglement entropy, which can be computed from the spectrum of the truncated correlation matrix.
\\

\noindent\textbf{Entanglement entropy and truncated correlation matrix. } Let us consider the ground state $|\Psi_0\rrangle$ of the spin chain Hamiltonian $\mathcal{H}$, which is obtained by filling the Fermi sea,
\begin{equation}\label{eq:GS}
    |\Psi_0\rrangle = \tilde{c}_0^\dagger \tilde{c}_1^\dagger \tilde{c}_2^\dagger \dots \tilde{c}_K^\dagger  |0\rrangle,
\end{equation}
where it is assumed that the energies are ordered such that $\omega_k < 0$ if and only if $k \leqslant K$. We are interested in the entanglement between a subsystem $A$, composed of the first $\ell + 1$ sites of the chain, and its complement $B$. This entanglement is characterized by the entanglement entropy $S$, defined as the von Neumann entropy of the reduced density matrix $\rho_A$ associated with region $A$,
\begin{equation}
    S = - \text{Tr}_A \left( \rho_A \ln \rho_A \right),
\end{equation}
with $\rho_A$ obtained by taking the partial trace with respect to the degrees of freedom of $B$,
\begin{equation}
    \rho_A = \text{Tr}_B |\Psi_0\rrangle \llangle \Psi_0 |.
\end{equation}
It follows from Wick's theorem that this matrix can be expressed as a thermal state of a so-called entanglement Hamiltonian $\mathcal{H}_{\rm ent}$ \cite{peschel2003calculation}, i.e.
\begin{equation}
    \rho_A = \frac{e^{- \mathcal{H}_{\rm ent}}}{\mathcal{Z}}, \quad \mathcal{Z} = \text{Tr}_A \left(e^{- \mathcal{H}_{\text{ent}}} \right),
\end{equation}
where $\mathcal{H}_{\text{ent}}$ takes the form of a free fermion Hamiltonian defined exclusively on region $A$,
\begin{equation}\label{eq:ent-hamil}
    \mathcal{H}_{\rm ent} = \sum_{m,n \in A} h_{mn} c_m^\dagger c_n.
\end{equation}
The matrix $h$ whose entries arise in Eq.~\eqref{eq:ent-hamil} was shown in \cite{peschel2003calculation} to be given in terms of the $(\ell+1) \times (\ell+1) $ submatrix $C$ of the correlation matrix as
\begin{equation}\label{eq:relhC}
    h = \ln\left((1-C)/C\right).
\end{equation}
The entries $C_{mn}$ of $C$, with $m, n \in \{0,1,\dots, \ell\}$, are the two-point correlations,
\begin{equation}\label{eq:Cmn}
    C_{mn} = \llangle \Psi_0 | c_m^\dagger c_n |\Psi_0\rrangle = \sum_{k =0}^K \phi_m(\omega_k)\phi_n(\omega_k).
\end{equation}
Using the relationship between $\rho_A$, $h$, and $C$, the entanglement entropy $S$ can be expressed in terms of the eigenvalues of this truncated correlation matrix $C$ \cite{peschel2009reduced}. Denoting the eigenvalues and eigenvectors of $C$ by $\gamma_n$ and $\ket{\gamma_n}$, respectively, i.e.
\begin{equation}
    C \ket{ \gamma_n} = \gamma_n \ket{\gamma_n}, \quad n = 0,1,\dots, \ell,
\end{equation}
the entanglement entropy $S$ is then given by
\begin{equation}\label{eq:entent}
    S = - \sum_{n=0}^\ell \left( \gamma_n \ln \gamma_n + (1 - \gamma_n) \ln (1 - \gamma_n) \right).
\end{equation}
This formula can be used to investigate analytically or numerically the behavior of the entanglement, particularly its scaling with the system size $N$. Since inhomogeneous free fermion models are critical, the entanglement entropy is expected to scale according to an area law with logarithmic corrections {\cite{VLRK03,CC04}. For a one-dimensional chain, where subsystem $A$ consists of the first $\ell + 1$ sites, this scaling takes the form
\begin{equation}\label{eq:scaling_cft}
    S \propto \ln(\ell + 1),
\end{equation}
in the limit $N \rightarrow \infty$ with a fixed ratio $(\ell+1) / (N+1)$. This result can be derived using conformal field theory (CFT) arguments, as inhomogeneous XX spin chains are effectively described by CFT in curved spacetimes in the large $N$ limit \cite{DSVC17,finkel2021entanglement}. This scaling behavior was also observed for the Krawtchouk chain, where the parameter $p$ governing the chain's inhomogeneities appears only in the subleading terms. Those subleading terms in fact display inhomogeneity-dependent oscillations, which is an additional peculiar feature of inhomogeneous chains \cite{bernard2022entanglement,bernard2024entanglement}. This is illustrated in Fig.~\ref{fig:Kraw_entanglement}, which depicts the entanglement entropy $S$ of Krawtchouk chains as a function of $N$, for different values of $p$ and for fixed filling fraction $(K+1) /(N+1)$ and aspect ratio $(\ell+1) /(N+1)$. 

\begin{figure}[h]
    \centering
\includegraphics[width=0.7\linewidth]{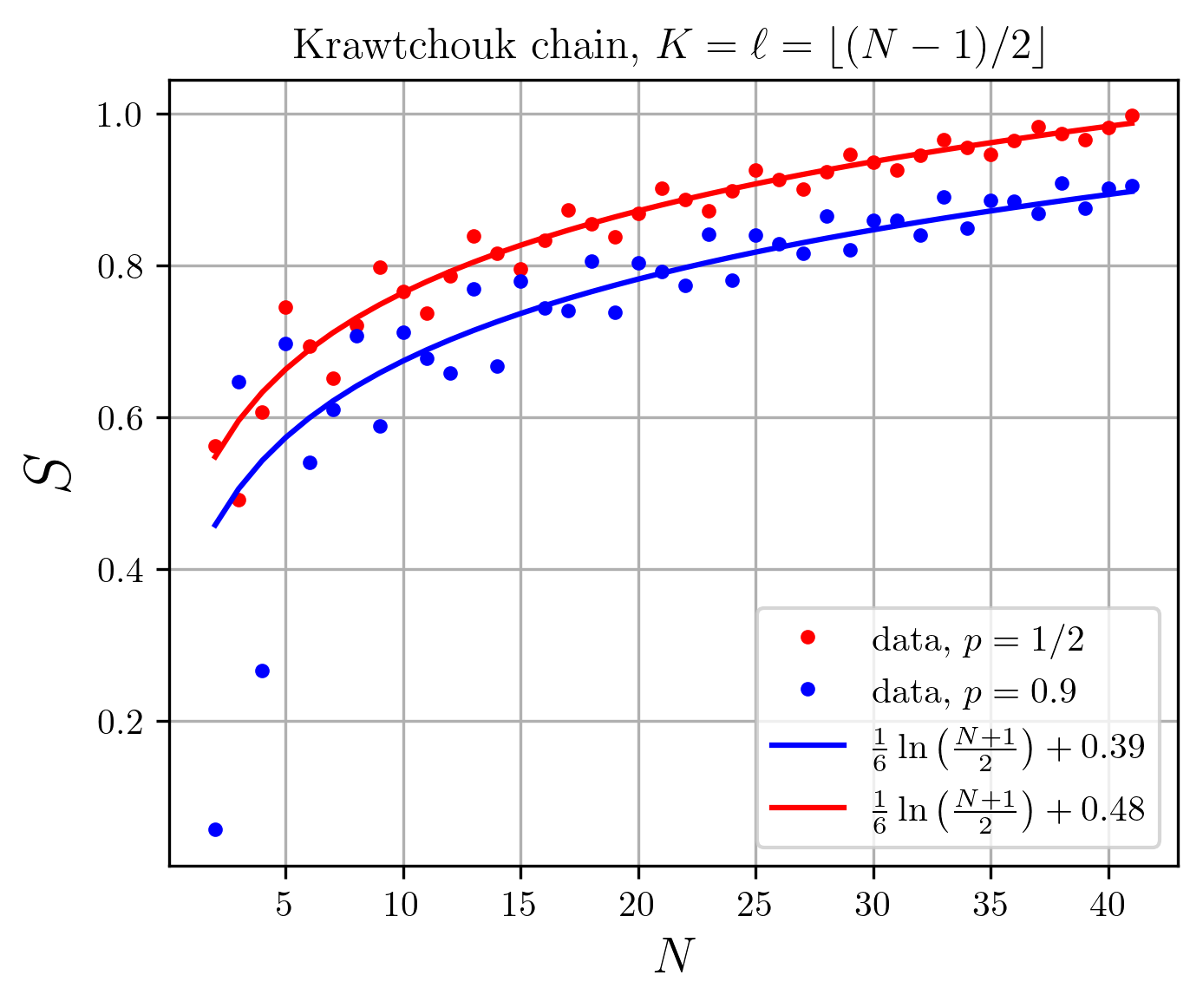}
    \caption{Entanglement entropy $S$ of Krawtchouk chains $p = 1/2$ and $p = 0.9$ as a function of $N$, with fixed filling fraction $(K+1) /(N+1)$ and aspect ratio $(\ell+1) /(N+1)$. The data originates from numerical diagonalization of the truncated correlation matrix. The two lines capture the scaling \eqref{eq:scaling_cft} predicted by CFT and constant subleading terms obtained by fitting the data. The presence in the data of subleading oscillations is apparent. }
    \label{fig:Kraw_entanglement}
\end{figure}

\noindent \textbf{Commuting tridiagonal matrix. } It was first observed in \cite{peschel2004reduced} that there are situations where a simple tridiagonal matrix that commutes with the truncated correlation matrix $C$ can be found. Such a commuting matrix was later shown to exist for the general case of inhomogeneous spin chains based on the bispectral orthogonal polynomials $\chi_n(x)$ of the Askey scheme \cite{crampe2019free,crampe2021entanglement}. In this context, one can leverage the bispectral relations of the polynomials to identify a simple tridiagonal matrix $T$ that commutes with the truncated correlation matrix $C$, thereby sharing a common eigenbasis. This goes as follows.

The polynomials of the discrete part of the Askey scheme correspond one-to-one with orthogonal polynomials that satisfy both a three-term recurrence relation and a three-term difference relation. This relationship allows for the identification of a pair of matrices, $\Lambda$ and $X$, whose actions on each other's eigenbasis is tridiagonal,
\begin{equation}\label{eq:rel1}
    \Lambda \ket{\omega_k} = \omega_k \ket{\omega_k}, \quad \Lambda \ket{\ell} = J_{\ell} \ket{\ell+1} + B_\ell \ket{\ell} + J_{\ell-1} \ket{\ell - 1}
\end{equation}
\begin{equation}\label{eq:rel2}
    X \ket{\omega_k} = \Bar{J}_{k} \ket{\omega_{k+1}} +  \Bar{B}_{k}  \ket{\omega_{k}} +  \Bar{J}_{k-1} \ket{\omega_{k-1}}, \quad X \ket{\ell} = \lambda_\ell \ket{\ell}.
\end{equation}
The matrix $\Lambda$ serves to define the Hamiltonian of the spin chain and represents its $1$-excitation Hamiltonian, while the matrix $X$ corresponds to the position operator for $1$-excitation states. Using relations \eqref{eq:rel1} and \eqref{eq:rel2}, one can show that the following $(\ell + 1) \times (\ell+1)$ matrix $T$ satisfies $[T,C] = 0$,
\begin{equation}\label{eq:deft}
   T = |\widehat{T}_{mn}|_{m,n \in A}, \quad \widehat{T} = \{ \Lambda - \omega_{K + 1/2}, X - \lambda_{\ell + 1/2} \},
\end{equation}
where $ \omega_{K + 1/2} := \frac{\omega_K +  \omega_{K+1}}{2}$ and $ \lambda_{\ell + 1/2} := \frac{\lambda_\ell + \lambda_{\ell + 1}}{2}$. This matrix $T$ is referred to as an \textit{algebraic Heun operator} because the Heun operator that defines the standard ODE with four regular singularities is obtained by repeating this construction with the bispectral operators of the Jacobi polynomials \cite{grunbaum2018algebraic}. The entries of the Heun operator $T$ in the eigenbasis of $X$ are given by
\begin{equation}
T=\begin{pmatrix}
\label{eq:Ttrid}
d_0 & t_0 & & & \\
t_0 & d_1 & t_1 & & \\
 & t_1 & d_2 & t_2 & \\
  & & \ddots & \ddots& \ddots\\
\end{pmatrix}
\end{equation}
where $t_n$ and $d_n$ are the following coefficients,
\begin{equation}
    t_n = J_n ( \lambda_n + \lambda_{n+1} - \lambda_{\ell} + \lambda_{\ell+1} ),
\end{equation}
\begin{equation}
    d_n = 2(B_n  - \omega_{K + 1/2})(\lambda_n -\lambda_{\ell + 1/2}).
\end{equation}

This construction is a reincarnation of the results of Slepian, Pollack, and Landau on time and band limiting problems \cite{slepian1961prolate}. Their initial analysis focused on the optimal way to concentrate in a finite time interval a signal limited to a definite frequency band. The solution of this problems requires the diagonalization of an integral operator. Their key observation was that a simple second-order differential operator commutes with this operator and is diagonalized by prolate spheroidal wavefunctions. In our case, the analog of limiting the signal in time is the restriction of the system to the region $A$. The restriction of the signal to a band of frequencies is akin to picking the subset of modes with negative energy that compose the Fermi sea and are excited in the ground state. The time and band limiting operators and their commuting differential operator are here associated with the truncated correlation $C$ matrix and its commuting $T$ matrix.

The identification of this commuting operator $T$ offers several benefits in the context of entanglement studies:

\begin{enumerate}
    \item \textit{Stability of numerical diagonalization}. The matrix $C$ is ill-conditioned, with multiple eigenvalues concentrated near $0$ and $1$. This is due to the fact that $C$ can be expressed as a product of projection operators. This makes identifying its eigenvectors sensitive to errors and thus unstable when done numerically with standard precision. The tridiagonal matrix $T$, however, does not generally suffer from this issue and exhibits a well-spaced spectrum. To improve accuracy and numerical stability, one can obtain the spectrum of $C$ by first diagonalizing $T$ and then applying $C$ on their shared eigenvectors to find $\gamma_n$.\\

    \item \textit{Connection with the Bethe Ansatz}. The matrix $T$ defined in \eqref{eq:deft} is known to arise in transfer matrices associated with solutions to the reflection equation. This establishes a connection with the theory of integrable models and notably enables the use of Bethe Ansatz techniques. It was shown in \cite{bernard2023computation} that $T$ and $C$ can be diagonalized via the algebraic Bethe Ansatz, with their eigenvalues expressed in terms of the roots of certain Bethe equations.\\

    \item \textit{Approximation of the entanglement Hamiltonian $h$.} Since $T$ is non-degenerate and commutes with $h$, there exists a polynomial $P$ such that $h = P(T)$. Meanwhile, the entanglement entropy of the ground state of a local free-fermion model is known to scale with the system size $N$ according to an area law up to logarithmic corrections. This scaling underscores the local nature of interactions within the ground state, as most entanglement arises from correlations between degrees of freedom near the boundary of regions $A$ and $B$. Given this locality, it is expected that the reduced density matrix associated with region $A$ resembles a thermal state of a local Hamiltonian $\mathcal{H}_{\text{ent}}$. Consequently, the matrix $h$ in the definition of $\mathcal{H}_{\text{ent}}$ should primarily describe local couplings and should therefore be well-approximated by a tridiagonal matrix,
\begin{equation}
    h \approx \alpha_0 + \alpha_1 T.
\end{equation}
This approximation and its validity are discussed in more detail in the following section.
\end{enumerate}

A generalization of the commuting tridiagonal matrix was also identified in models of free fermions on distance-regular graphs like the Hadamard graph \cite{crampe2020entanglement}, the Hamming graphs \cite{bernard2023entanglement}, the Johnson graph \cite{bernard2023entanglementJ}, and the folded cube \cite{bernard2023absence}. A generalization was also shown to exist for fermions on inhomogeneous hyperplane lattices based on multivariate Krawtchouk polynomials \cite{bernard2022entanglement}. 

\section{Entanglement Hamiltonians and Heun operators}\label{sec:6}
Since the hopping matrix $h$ is expressed in terms of the truncated correlation matrix $C$ which commutes with $T$, it follows that $[h,T] =0$ and that
\begin{equation}
    h \ket{\gamma_n} = \varepsilon_n \ket{\gamma_n}, \quad T\ket{\gamma_n} = t_n \ket{\gamma_n},
\end{equation}
where $\ket{\gamma_n}$ is a common eigenvector of $C$, $T$, and $h$. In the previous section, we argued that the hopping matrix $h$ of the entanglement Hamiltonian can be well-approximated by an affine transformation of the Heun operator $T$, i.e.
\begin{equation}
    \varepsilon_n \approx \alpha_0 + \alpha_1 t_n.
\end{equation}
Here, we outline a scheme for determining the coefficients $\alpha_0$ and $\alpha_1$ in this approximation and demonstrate numerically that it provides an accurate representation of the entire entanglement spectrum for the Krawtchouk chain. This approach has also been successfully applied to the anti-Krawtchouk chain \cite{bernard2023absence} and the gradient chain \cite{bonsignori2024entanglement}. \\

\noindent\textbf{Approximation scheme.} Given any coefficients $ \alpha_0 $ and $ \alpha_1 $, one can define a \textit{Heun Hamiltonian} and \textit{Heun density matrix} as the following,
\begin{equation}
    \rho_{T}(\alpha_0, \alpha_1) = \frac{e^{-\mathcal{H}_T(\alpha_0, \alpha_1)}}{\text{Tr}_A(e^{-\mathcal{H}_T(\alpha_0, \alpha_1)})}, \quad \mathcal{H}_T(\alpha_0, \alpha_1) = \sum_{m,n \in A} (\alpha_0 + \alpha_1 T)_{mn} c_m^\dagger c_n.
\end{equation}
One then expects $\rho_{T}(\alpha_0, \alpha_1)$ to provide a good approximation of the reduced density matrix $\rho_A$ when the matrix $\alpha_0 + \alpha_1 T$ is close to the hopping matrix $h$. In particular, when this is the case both $\rho_{T}(\alpha_0, \alpha_1)$ and $\rho_A$ should lead to similar expectation values for most observables. Good candidates for $\alpha_0$ and $\alpha_1$ can thus be selected by solving the following conditions ensuring that the two density matrices lead to the same entanglement entropy and the same expected number of excitation:
\begin{equation}\label{eq:condapp1}
    \langle Q_A \rangle =  \text{Tr}_A(Q_A\rho_A) = \text{Tr}_A(Q_A\rho_T(\alpha_0, \alpha_1)),
\end{equation}
\begin{equation}\label{eq:condapp2}
  S=   -\text{Tr}_A(\rho_A \ln \rho_A) = -\text{Tr}_A(\rho_T(\alpha_0, \alpha_1) \ln \rho_T(\alpha_0, \alpha_1)),
\end{equation}
where $Q_A = \sum_{n \in A} \frac{1+\sigma_n^z }{2}$. Since both the expected number of excitations $ \langle Q_A \rangle$ and the entanglement entropy $S$ can be expressed in terms of the truncated correlation matrix, it is straightforward to find numerically the parameters $\alpha_0$ and $\alpha_1$ satisfying these conditions. Once obtained, one can check the accuracy of the approximation by comparing the entanglement spectrum (i.e. the spectrum of $h$) obtained from $C$ using Eq.~\eqref{eq:relhC} to that of the hopping matrix $\alpha_0 + \alpha_1 T$.  The comparison of $\varepsilon_n$ and $\alpha_0 + \alpha_1 t_n$ is provided in the case of the symmetric Krawtchouk chain in Fig.~\eqref{fig:Kraw}. It illustrates that this approximation scheme can indeed provide a tridiagonal matrix that commutes with $h$ and qualitatively captures its spectrum. To obtain a more quantitative comparison between the affine approximation and the exact entanglement Hamiltonian, one can further study the R\'enyi fidelities between the respective density matrices \cite{parez2022symmetry,bernard2023absence}. \\
\begin{figure}
    \centering
    \includegraphics[width=0.7\linewidth]{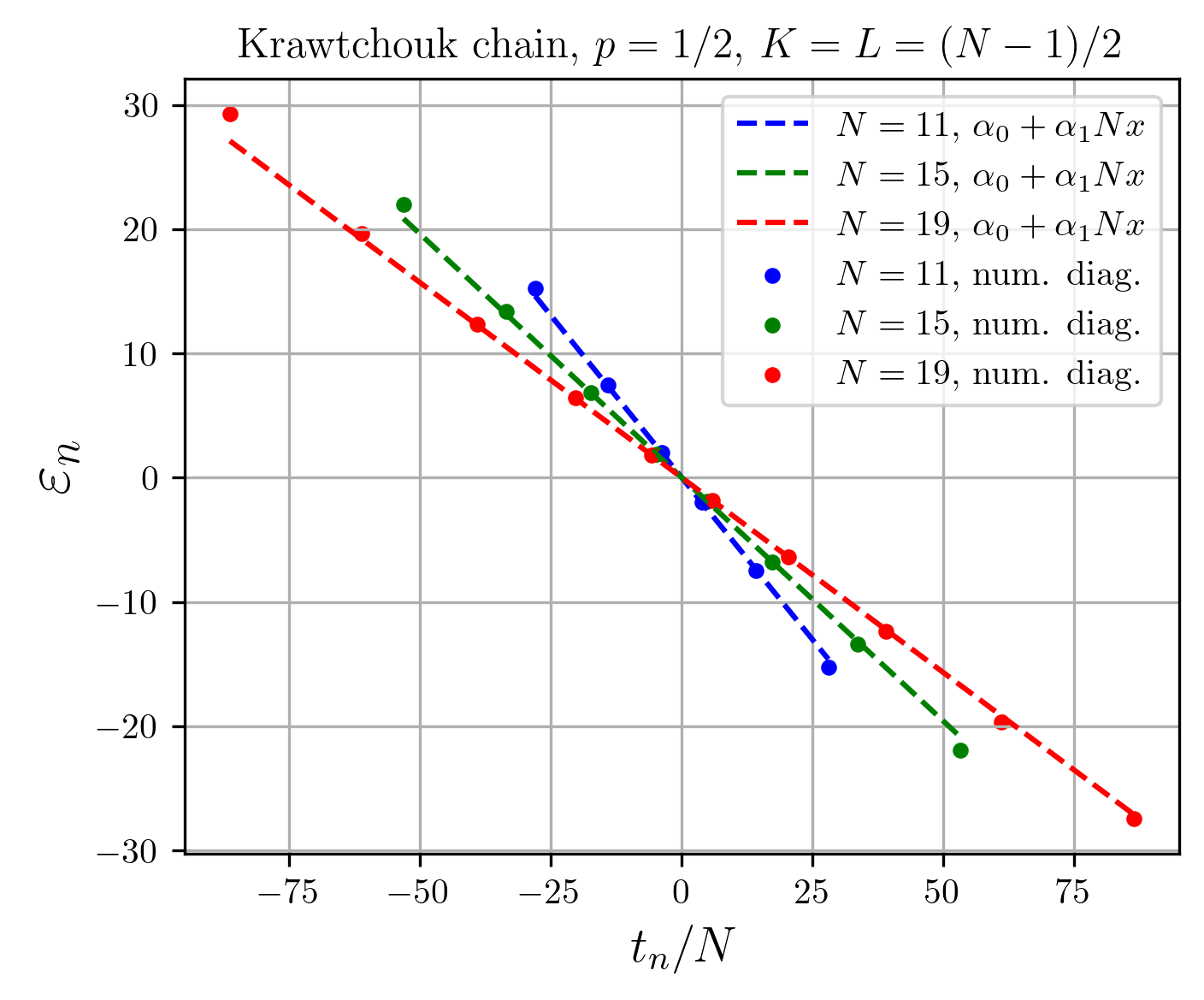}
    \caption{Comparison of the entanglement spectrum $\varepsilon_n$ obtained by numerical diagonalization of $h$ computed from the truncated correlation matrix to the spectrum of $\alpha_0 + \alpha_1 T$ with $\alpha_0$ and $\alpha_1$ obtained by solving equations \eqref{eq:condapp1} and \eqref{eq:condapp2}, for the symmetric Krawtchouk chain.}
    \label{fig:Kraw}
\end{figure}

\noindent \textbf{CFT prediction. } A further motivation for the approximation $h \approx \alpha_0 + \alpha_1 T$ was put forward by Bonsignori and Eisler, using predictions from CFT. The large-$N$ limit of inhomogeneous XX spin chains can be effectively modeled by a CFT defined on a curved spacetime as already mentioned. Using this connection, they were able to identify a discretization of the entanglement Hamiltonian provided by CFT predictions. This discretization was shown to fit the large-$N$ limit of Heun operators for models like the gradient chain. This observation will be extended to spin chains based on Racah polynomials and their associated Heun-Racah operator in an upcoming work \cite{upcom}.

\section{Entanglement negativity and correlations}\label{sec:7}

The entanglement entropy and entanglement Hamiltonian discussed in previous sections capture quantum correlations between complementary regions $A$ and $B$ of a quantum many-body system in a pure state. However, these measures do not properly quantify entanglement in mixed states, or between non-complementary regions. An important example is that of a tripartite system $A_1\cup A_2 \cup B$, where the non-complementary regions $A_1$ and $A_2$, of respective lengths $\ell_1$ and $\ell_2$, are separated by a distance $d$. In particular, one is typically interested in the decay of entanglement with the separation between the regions. In this context, one uses the \textit{logarithmic negativity} $\mathcal{E}_{b/f}$ \cite{VW02,plenio2005logarithmic,SSR17}, where the indices $b,f$ stand for bosonic and fermionic, respectively. Indeed, the definition of the logarithmic negativity depends on the statistics of the particles involved. 
In the following, we review recent results \cite{blanchet24Neg} regarding the fermionic logarithmic negativity in inhomogeneous free-fermion models described by the Hamiltonian \eqref{eq:defHamil}.\\

\noindent \textbf{Fermionic logarithmic negativity and the correlation matrix. } 
The fermionic logarithmic negativity is defined as \cite{SSR17}
\begin{equation}
 \mathcal{E}_f = \ln \big \lVert \rho_{A_1,A_2}^{R_1}\big \rVert_1,
\end{equation}
where $\rho_{A_1,A_2}$ is the reduced density matrix of the system $A_1\cup A_2$$, \lVert X \rVert_1 = \Tr \sqrt{X X^\dagger}$ is the trace norm, and $R_1$ indicates the partial time reversal on~$A_1$. In the occupation basis, this operation is defined as follows. Consider basis states $|\alpha_1\beta_2\rrangle$  for $A_1 \cup A_2$ of the form 
\begin{equation}
    |\alpha_1\beta_2\rrangle = \prod_{j\in A_1} (c_j^\dagger)^{n_j} \prod_{j'\in A_2} (c_{j'}^\dagger)^{n_{j'}}|0\rrangle, 
\end{equation}
with $n_{j}\in \{0,1\}$. 
The partial time reversal operation is then defined as \cite{SSR17}
\begin{equation}
    \Big( |\alpha_1\beta_2\rrangle\, \llangle\widetilde\alpha_1\widetilde\beta_2|\Big)^{R_1}=(-1)^{\phi(\{n_j\},\{\tilde n_j\})}|\widetilde\alpha_1\beta_2\rrangle\, \llangle\alpha_1\widetilde\beta_2|
\end{equation}
where the phase is 
\begin{equation}
   \phi(\{n_j\},\{\tilde n_j\}) = \frac{f_1(f_1+2)}{2}+ \frac{\tilde f_1(\tilde f_1+2)}{2} + f_2 \tilde f_2 + f_1 f_2 + \tilde f_1 \tilde f_2 + (f_1+f_2)(\tilde f_1 + \tilde f_2)
\end{equation}
and $f_{1,2},\tilde f_{1,2}$ are the local occupation numbers in the basis states for $A_1$ and $A_2$. In free-fermion models, the fermionic logarithmic negativity can be obtained from the truncated correlation matrix $C$ pertaining to $A_1\cup A_2$, similarly to the entanglement entropy. For simplicity, we introduce the covariance matrix $J = 2 C - \mathbb{I}$. Because $A_1$ and $A_2$ are disjoint, the matrix has a block structure,
\begin{equation}
        J = \begin{pmatrix}
        J_{11} & J_{12} \\
        J_{21} & J_{22}\\
    \end{pmatrix},
    \label{eq:J Block}
\end{equation}
where $J_{ij}$ is an $\ell_i \times \ell_j$ matrix which encodes the correlations between $A_i$ and $A_j$. To proceed, we introduce the following matrices,
\begin{equation}
\begin{split}
    J_{\pm}& = \begin{pmatrix}
        -J_{11} & \pm i J_{12} \\
        \pm i J_{21} & J_{22}\\
    \end{pmatrix}, \\[.3cm]
     J_{\rm x} &= (\mathbb{I} + J_+ J_-) (J_+ + J_-),
    \end{split}
    \label{eq:Jpm}
\end{equation}
and the fermionic logarithmic negativity reads \cite{SR19,SRRC19}
\begin{equation}
    \mathcal{E}_f = \Tr_{A_1\cup A_2} \ln \left[ \left( \frac{\mathbb{I}+J_{\rm x}}{2} \right)^{\frac{1}{2}} + \left( \frac{\mathbb{I}-J_{\rm x}}{2} \right)^{\frac{1}{2}} \right] + \frac{1}{2} \Tr_{A_1\cup A_2} \ln \left[ \left( \frac{\mathbb{I}+J}{2} \right)^2 + \left( 
 \frac{\mathbb{I}-J}{2}\right)^2 \right].
\label{eq:LN}
\end{equation}

\noindent \textbf{Negativity of adjacent regions in the Krawtchouk chain. } For simplicity, we focus on the Krawtchouk chain, where the couplings are given in Eq.~\eqref{eq:coeff_kraw}. We study the logarithmic negativity of adjacent regions in the bulk of the chain (far from the boundaries) in the ground state given in Eq.~\eqref{eq:GS}. At half filling we find the scaling \cite{blanchet24Neg}
\begin{equation}
    \mathcal{E}_{f}= \frac{1}{4} \log \Big(\frac{\ell_1 \ell_2}{\ell_1+\ell_2}\Big)+\textrm{cst}
\end{equation}
in the limit $N\to \infty$ with fixed ratios $\ell_i/N$. This scaling corresponds to the CFT prediction for a theory with central charge $c=1$ \cite{CCT12,CCT13,SSR17}, and this result is coherent with previous entanglement studies in the Krawtchouk chain \cite{finkel2021entanglement,bernard2022entanglement}. \\

\noindent \textbf{Skeletal regime and bulk/boundary negativity. } To examine how the fermionic logarithmic negativity depends on the separation $d$ between disjoint regions, we employ the skeletal regime \cite{BWK22}, which considers the case where both regions are reduced to a single site, i.e., $\ell_1=\ell_2=1$. This regime is sufficient to extract the leading terms in the entanglement decay, and has been applied successfully in the context of Dirac fermions in arbitrary dimensions \cite{parez2024entanglement}, the Schwinger model \cite{florio2024two} and the Krawtchouk chain \cite{blanchet24Neg}. In this regime, the fermionic logarithmic negativity between sites $m$ and $n$ scales at leading order as \cite{parez2024entanglement,blanchet24Neg}
\begin{equation}
    \mathcal{E}_f = \frac{2}{1+2(\rho-1)\rho} |C_{mn}|^2 
    \label{eq:EfC}
\end{equation}
in the limit of large separation $d=|m-n|$ and system size. The function $C_{mn}$ is the two-point correlation function defined in \eqref{eq:Cmn} and $\rho$ is the filling fraction, defined as $\rho=\frac{K+1}{N+1}$. Combining analytical and numerical calculations, we obtain the following results for the Krawtchouk chain \cite{blanchet24Neg}: \\

\begin{enumerate}
    \item {\it Bulk power-law decay.} For $m=pN-d/2$ and $n=pN+d/2$ with $d\ll N$ and $N\to \infty$ (deep in the bulk), we find 
    \begin{equation}
    C_{pN-\frac d2,pN+\frac d2 } \sim \frac{1}{\pi d}\sin\left( d \sqrt{\frac{1}{p(1-p)}}  \arcsin (\sqrt{\rho})\right)
    \label{eq:CGenCOnj}
\end{equation}
    which corresponds to the following power-law decay for the negativity,
    \begin{equation}
        \mathcal{E}_f\propto d^{-4 \Delta_f}, \quad \Delta_f=1/2.
    \end{equation}
    This is the same behavior as free Dirac fermions in one dimension \cite{parez2024entanglement}. \\
    
\item {\it Boundary power-law decay.} For $m$ at, or close to, the left boundary, we have the following analytical results in the large-$N$ limit for $\rho=p$,
\begin{align}
    C_{0,d} & \sim\frac{1}{(2\pi^3)^{\frac 14} d^{\frac 34}}\sin\Big(\frac{-\pi d}{2}\Big), \label{eq:C0d} \\
    C_{1,d+1} & \sim \frac{1}{(2\pi^3)^{\frac 14} d^{\frac 54}}\sin\Big(\frac{-\pi d}{2}\Big). \label{eq:C1dplus1}
\end{align}
    Combining these results with numerical ones, we find that the boundary negativity decays as a power-law, $\mathcal{E}_f\propto d^{-4 \Delta_f},$ where $\Delta_f$ depends on the parity of the leftmost site $m=0,1,2,3,\dots$, i.e.
    \begin{equation}
        \Delta_f^{\textrm{even}}=3/8, \quad \Delta_f^{\textrm{odd}}=5/8. 
    \end{equation}
    In the homogeneous chain, the boundary decay is the same as in the bulk, namely $\mathcal{E}_f\propto d^{-2}$. The parity dependence observed here is thus a striking property caused by the inhomogeneity of the model. 
\end{enumerate}
We conclude this section by mentioning that the power-law decays of entanglement observed here hold in the fermionic picture of the model. Since the Jordan-Wigner transformation is non-local in terms of the spins, reduced density matrices for disjoint spins do not match the corresponding fermionic ones \cite{coser2015partial}. In related inhomogeneous spin models described by the Hamiltonian \eqref{eq:hamil}, we instead expect the negativity between distant spins to vanish beyond a critical separation. Indeed, the \textit{fate of entanglement} picture \cite{parez2024fate} predicts that entanglement between disjoint regions in (finite) spin/bosonic systems should suffer an entanglement sudden death at finite separation, whereas for fermionic ones the parity superselection rule precludes the sudden death, and one typically has a power-law decay of entanglement. 
This extremely general result has notably been observed in highly-entangled states, such as resonating valence-bond states \cite{parez2023separability} and quantum critical ground~states~\cite{OAFF02,JTBBJH18,parez2024entanglement}.

\section{Outlook}

We reviewed various results concerning the study of inhomogeneous XX spin chains, or equivalently, inhomogeneous free-fermion chains. The standard diagonalization approach for this model was recalled, with connections to orthogonal polynomials highlighted. We discussed how these spin chains can serve as models for quantum communication channels when they exhibit PST, and we reviewed the conditions required on inhomogeneous couplings for this phenomenon. Recent results on heat currents in these spin chains, when coupled to thermal baths at different temperatures, were also discussed. In particular, the relation to the mirror symmetry condition needed for PST was underscored. We also reviewed significant findings regarding ground state entanglement in these models. We recalled the standard approach for computing entanglement entropy and emphasized the role of a commuting Heun operator in chains based on polynomials of the Askey-scheme, arguing that this operator could serve as an effective approximation for the hopping matrix in the entanglement Hamiltonian. {Finally, we reviewed recent results regarding the entanglement negativity and the unusual decay of boundary correlations in the Krawtchouk chain.}

Despite their simplicity, XX spin chains model a wide array of phenomena and allow for the derivation of many analytical and numerical results, owing to their connection with the theory of orthogonal and special functions. Several open questions remain regarding these models, which we aim to investigate in future work. Among these is the identification of the thermodynamic properties of spin chains that could be tied to the spectral conditions required for PST. Additionally, we plan to further explore the connection between the hopping matrix and the Heun operator in the Krawtchouk chain, with the goal of deriving an analytic approximation of $h$ as an affine transformation of $T$. {Another promising avenue would be to delve further into the multipartite entanglement structure of these models.}

\section*{Acknowledgments}
PAB holds an Alexander-Graham-Bell scholarship from the Natural Sciences and Engineering Research Council (NSERC) of Canada. LV is funded in part through a discovery grant from NSERC.

\providecommand{\href}[2]{#2}\begingroup\raggedright\endgroup

\end{document}